\documentstyle[aps,pra]{revtex}
\begin{document}
\draft

\newcommand{\pp}[1]{\phantom{#1}}
\newcommand{\be}{\begin{eqnarray}}
\newcommand{\ee}{\end{eqnarray}}
\newcommand{\ve}{\varepsilon}
\newcommand{\vs}{\varsigma}
\newcommand{\Tr}{{\,\rm Tr\,}}
\newcommand{\pol}{\frac{1}{2}}

\title{
Notes on nonlinear quantum algorithms
}
\author{Marek Czachor}
\address{
Wydzia{\l}  Fizyki Technicznej i Matematyki Stosowanej\\
 Politechnika Gda\'{n}ska,
ul. Narutowicza 11/12, 80-952 Gda\'{n}sk, Poland
}
\maketitle

\begin{abstract}
Recenty Abrams and Lloyd \cite{AL} have proposed a fast algorithm that
is based on a nolinear evolution of a state of a quantum
computer. They have explicitly used the fact that nonlinear
evolutions in Hilbert spaces do not conserve scalar products of
states, and applied a description of separated systems taken from
Weinberg's nonlinear quantum mechanics. On the other hand it is
known that violation of orthogonality combined with the
Weinberg-type description generates unphysical, arbitrarily fast
influences between noninteracting systems. It was not therefore
clear whether the algorithm is fast because arbitrarily fast
unphysical effects are involved. 
In these notes I show that this is not the case. I analyze both
algorithms proposed by Abrams and Lloyd on concrete, simple
models of nonlinear evolution. The description I choose is known
to be free of the unphysical influences (therefore it is not the
Weinberg one). I show, in particular, that the correct local
formalism allows even to simplify the algorithm.
\end{abstract}

\section{First algorithm}

\noindent
{\it Step 1.\/} 
We begin with the state 
\be
|\psi[0]\rangle &=&|0_1,\dots,0_n\rangle|0\rangle
\ee
where the first $n$ qubits correspond to the input and the last
qubit represents the output.

Consider the unitary transformation acting as follows
\be
U|0\rangle &=&\frac{1}{\sqrt{2}}\Big(|0\rangle+ |1\rangle\Big)\\
U|1\rangle &=&\frac{1}{\sqrt{2}}\Big(-|0\rangle+ |1\rangle\Big)
\ee
{\it Step 2.\/} 
\be
|\psi[1]\rangle &=&\underbrace{U\otimes\dots\otimes U}_n\otimes 1
|\psi[0]\rangle\\
&=&\frac{1}{\sqrt{2^n}}\sum_{i_1\dots i_n=0}^1
|i_1,\dots,i_n\rangle|0\rangle
\ee
The input constists now of a uniform superposition of all the
numbers $0\leq n \leq 2^n-1$.

\medskip
\noindent
{\it Step 3.\/} 
\be
|\psi[2]\rangle &=& F|\psi[1]\rangle\\
&=&
\frac{1}{\sqrt{2^n}}\sum_{i_1\dots i_n=0}^1
|i_1,\dots,i_n\rangle|f(i_1,\dots,i_n)\rangle
\ee
where $F$ is some unitary transformation (oracle) that transforms
the input into an output; $f(i_1,\dots,i_n)$ equals 1 or 0.

\medskip
\noindent
{\it Step 4.\/} 
\be
|\psi[3]\rangle &=&\underbrace{U^{-1}\otimes\dots\otimes U^{-1}}_n\otimes 1
|\psi[2]\rangle\\
&=&\frac{1}{2^n}\sum_{i_1\dots i_n=0}^1
\Big(|0_1\rangle+(-1)^{i_1+1}|1_1\rangle\Big)\otimes
\dots \otimes
\Big(|0_n\rangle+(-1)^{i_n+1}|1_n\rangle\Big)\otimes|f(i_1,\dots,i_n)\rangle\\
&=&\frac{1}{2^n}\sum_{j_1\dots j_n=0}^1\sum_{i_1\dots i_n=0}^1
(-1)^{(i_1+1)j_1+\dots + (i_n+1)j_n}
|j_1,\dots,j_n\rangle
|f(i_1,\dots,i_n)\rangle\\
&=&\frac{1}{2^n}\sum_{i_1\dots i_n=0}^1
|0_1,\dots,0_n\rangle
|f(i_1,\dots,i_n)\rangle\nonumber\\
&\pp =&+\frac{1}{2^n}\sum_{\{j_1\dots j_n\}\neq
\{0_1\dots 0_n\}} \sum_{i_1\dots i_n=0}^1
(-1)^{(i_1+1)j_1+\dots + (i_n+1)j_n}
|j_1,\dots,j_n\rangle
|f(i_1,\dots,i_n)\rangle\\
&=&
|0_1,\dots,0_n\rangle\Big(\frac{2^n-s}{2^n}
|0\rangle+\frac{s}{2^n}|1\rangle\Big)\nonumber\\
&\pp =&+\frac{1}{2^n}\sum_{\{j_1\dots j_n\}\neq
\{0_1\dots 0_n\}} \sum_{i_1\dots i_n=0}^1
(-1)^{(i_1+1)j_1+\dots + (i_n+1)j_n}
|j_1,\dots,j_n\rangle
|f(i_1,\dots,i_n)\rangle\label{state}.
\ee
The probability of finding the input in the state
$|0_1,\dots,0_n\rangle$ is 
\be
P(s)=\frac{(2^n-s)^2+s^2}{2^{2n}}
\ee
$P(s)$ is a parabola satisfying $P(0)=P(2^n)=1$ which shows that
it has a minimum in $s=2^{n-1}$. The minimal probability of
finding the input in the state $|0_1,\dots,0_n\rangle$ is
therefore $P(2^{n-1})=1/2$ and it occurs if $s$ is  
exactly one-half of $2^n$. 

Probability of finding $f(i_1,\dots,i_n)=1$ is $s/2^n$. This
intuitively natural result becomes less natural if one tries to
prove it by using (\ref{state}). So let's do it explicitly. 
Let us begin with rewriting (\ref{state}) in the following form
\be
|\psi[3]\rangle 
&=&
\frac{1}{2^n}\sum_{j_1\dots j_n}
\sum_{\{i_1\dots i_n;\,f(i_1,\dots,i_n)=1\}}
(-1)^{(i_1+1)j_1+\dots + (i_n+1)j_n}
|j_1,\dots,j_n\rangle
|1\rangle\nonumber\\
&\pp =&+
\frac{1}{2^n}\sum_{j_1\dots j_n}
\sum_{\{i_1\dots i_n;\,f(i_1,\dots,i_n)=0\}}
(-1)^{(i_1+1)j_1+\dots + (i_n+1)j_n}
|j_1,\dots,j_n\rangle
|0\rangle
\ee
The probability of finding the flag qubit in $|1\rangle$ is
\be
P_f(1)
&=&
\frac{1}{4^n}\sum_{j_1\dots j_n}
\Big|\sum_{\{i_1\dots i_n;\,f(i_1,\dots,i_n)=1\}}
(-1)^{(i_1+1)j_1+\dots + (i_n+1)j_n}\Big|^2\\
&=&
\frac{1}{4^n}\sum_{j_1\dots j_n}
\Big|\sum_{\{i_1\dots i_n;\,f(i_1,\dots,i_n)=1\}}
(-1)^{i_1j_1+\dots + i_nj_n}\Big|^2\\
&=&
\frac{1}{4^n}\sum_{\vec j}
\Big|
(-1)^{\vec i^1\cdot\vec j}+\dots +
(-1)^{\vec i^s\cdot\vec j}
\Big|^2
\ee
where the vectors $\vec i^r$, $r=1\dots s$, are all
different (which is essential for the proof) and $\vec j=(j_1,\dots,j_n)$. 
\be
P_f(1)
&=&
\frac{1}{4^n}\sum_{\vec j}
\Big(s +\sum_{k\neq l}
(-1)^{(\vec i^k+\vec i^l)\cdot\vec j}
\Big)=
\frac{1}{4^n}
\Big(s2^n +\sum_{k\neq l}\sum_{\vec j}
(-1)^{(\vec i^k+\vec i^l)\cdot\vec j}
\Big)=\frac{s}{2^n}
\ee
because the sum over $\vec j$ vanishes.

\medskip
\noindent
{\it Step 5.\/} 
We want to distinguish between the cases $s=0$ and $s>0$ for
small $s$. To do so we are going to use a nonlinear dynamics
that does not change the ``North Pole" $|0\rangle$ but any
superposition of $|0\rangle$ with $|1\rangle$ drags to the
``South". The violation of orthogonality is called, after
Mielnik \cite{Mielnik}, the mobility phenomenon.

\section{Mobility frequency}

Let us first concentrate on a single-qubit system. The first
natural guess is something like
\be
i|\dot \psi\rangle &=& \epsilon\Big(
\frac{\langle\psi|A|\psi\rangle}{\langle\psi|\psi\rangle}
-
\langle 0|A|0\rangle\Big)A |\psi\rangle\label{S1}
\ee
where 
\be
A=\eta \Big(|0\rangle\langle 0|
- |1\rangle\langle 1|\Big)
+
\sqrt{1-\eta^2}\Big(|0\rangle\langle 1|+|1\rangle\langle
0|\Big) 
\ee
and $\eta$ is small but nonzero. The solution of (\ref{S1}) for
normalized $\psi_0$ is 
\be
|\psi_t\rangle
&=&
\exp\Big[-i\epsilon\Big(
\langle\psi_0|A|\psi_0\rangle
-
\langle 0|A|0\rangle\Big)A t\Big]|\psi_0\rangle\\
&=&
\bbox 1\cos\Big[\epsilon\Big(
\langle\psi_0|A|\psi_0\rangle
-
\langle 0|A|0\rangle\Big) t\Big]|\psi_0\rangle
-iA
\sin\Big[\epsilon\Big(
\langle\psi_0|A|\psi_0\rangle
-
\langle 0|A|0\rangle\Big) t\Big]|\psi_0\rangle
\ee
Assume
\be
|\psi_0\rangle &=& 
\frac{2^n-s}{\sqrt{(2^n-s)^2+s^2}}
|0\rangle+\frac{s}{\sqrt{(2^n-s)^2+s^2}}|1\rangle
\ee
Then
\be
\langle\psi_0|A|\psi_0\rangle 
&=& 
\frac{1}{(2^n-s)^2+s^2}
\Big[(2^n-s)\langle 0|+s\langle 1|\Big]A
\Big[(2^n-s)|0\rangle+s|1\rangle\Big]\\
&=& 
\frac{(2^n-s)^2\eta-s^2\eta+2(2^n-s)s\sqrt{1-\eta^2}}{(2^n-s)^2+s^2}
\ee
The mobility frequency is therefore 
\be 
\omega_\epsilon=\epsilon\frac{(2^n-s)^2\eta-s^2\eta+2(2^n-s)s\sqrt{1-\eta^2}
-\eta(2^n-s)^2-\eta s^2}{(2^n-s)^2+s^2}
=
\epsilon\frac{-2s^2\eta+2(2^n-s)s\sqrt{1-\eta^2}}{(2^n-s)^2+s^2}
\ee
which for $2^n\gg s$ gives approximately 
\be
\omega_\epsilon\approx\epsilon\frac{s\sqrt{1-\eta^2}}{2^{n-1}}
\approx\frac{\epsilon s}{2^{n-1}}
\ee
which makes the algorithm exponentially slow.

Let us try therefore another nonlinearity:
\be
i|\dot \psi\rangle &=& \epsilon\tanh \Big(
\frac{\langle\psi|\psi\rangle}{\langle\psi|A|\psi\rangle}
-
\frac{1}{\langle 0|A|0\rangle}\Big)A |\psi\rangle
\ee
We find 
\be 
\omega'_\epsilon &=&
\epsilon\tanh \Big[
\frac{(2^n-s)^2+s^2}{(2^n-s)^2\eta-s^2\eta+2(2^n-s)s\sqrt{1-\eta^2}}
-
\frac{1}{\eta}\Big]\\
&=&
\epsilon\tanh\Big[
\frac{\eta(2^n-s)^2+\eta s^2
-(2^n-s)^2\eta+s^2\eta-2(2^n-s)s\sqrt{1-\eta^2}
}{(2^n-s)^2\eta^2-s^2\eta^2+(2^n-s)s\eta\sqrt{1-\eta^2}}\Big]\\
&=&
\epsilon\tanh\Big[
\frac{2\eta s^2-2(2^n-s)s\sqrt{1-\eta^2}
}{(2^n-s)^2\eta^2-s^2\eta^2+(2^n-s)s\eta\sqrt{1-\eta^2}}\Big]\\
&\approx &
\epsilon\tanh\Big[
\frac{-s\sqrt{1-\eta^2}
}{2^{n-1}\eta^2}\Big]\approx -\epsilon\tanh\Big[
\frac{s}{2^{n-1}\eta^2}\Big]
\ee
For $\eta$ of the order of $2^{-(n-1)/2}$ one can obtain a
reasonable mobility frequency but this requires an exponentially
precise control over $\langle 0|A|0\rangle$.  

\section{Evolution of the entire system}

The discussion given above applies to a single-qubit (flag)
subsystem. The entire system that is involved consists of $n+1$
systems and therefore we arrive at the delicate problem of
extending a one-particle nonlinear dynamics to more particles. 

The description chosen by Abrams and Lloyd uses the Weinberg
prescription. Several comments are in place here. First, it is known
that the Weinberg formulation implies a ``faster-than-light
telegraph". The version of the telegraph especially relevant in
this context is the one that is based on the mobility effect
\cite{MCfpl}. It is therefore not clear {\it a priori\/} to what
extent the fact that the algorithm is fast depends on the
presence of faster than light effects. Second, the Weinberg
prescription is meant to describe systems that do not interact.
We have two options now. Either we indeed want to keep the flag
qubit noninteracting with the input (during the nonlinear
evolution) or we allow a nonlinear evolution which involves the
entire quantum computer. If we decide on the first option we
should use the Polchinski-type description which eliminates the
unphysical nonlocal influences, but the nonlinear evolution of
the flag qubit is determined by its reduced density matrix (the
Polchinski-type description was recently formulated for
a class of equations more general than those considered by
Weinberg in \cite{MCqph}; its aplication to interacting systems
can be found for example in \cite{MCpra}). This
is the reduced density matrix obtained by the reduction over all
$2^n$ states of the input subsystem. 
Physically this kind of evolution occurs if the nonlinearity is
active independently of the state of the $n$ input qubits.

But the very idea of the algorithm is to take advantage of the fact
that probability of finding the entire input in the ground state
esceeds 1/4. It is also assumed that one can turn the
nonlinearity on and off. It is legitimate therefore to
contemplate the situation where the nonlinearity is turned on
only provided all the input detectors signal 0. 

At this point one might be tempted to act as follows: Take as an
initial condition for our nonlinear evolution the product state
obtained by projecting the entire entangled state on
$|0_1,\dots,0_n\rangle$. The problem with this kind of approach
is that the ``projection postulate" of linear quantum mechanics
does not have an immediate extension to a nonlinear dynamics.
There are many reasons for this but I do not want to discuss it here.
At this moment it is sufficient to know that it is safer to
avoid reasonings based on the projection postulate if
nonlinearity is involved. 

I propose an alternative formulation. Assume that indeed the
nonlinearity is activated only if the input is in the ground
state. In principle there is no problem with this because all the
different combinations of 0's and 1's correspond to orthogonal
vectors in the $2^n$-dimensionl Hilbert space of the input and
there exists, in principle, an analyzer that splits a beam of
input particles into $2^n$ different sub-beams. We can place our
hypothetical nonlinear medium in front of this output of the analyzer
that corresponds to the qubinary zero.

Let us introduce two projectors:
\be
P^{(n)}&=&|0_1,\dots,0_n\rangle\langle 0_1,\dots,0_n|\otimes
\bbox 1\\
P &=&\bbox 1^{(n)}\otimes|0\rangle\langle 0|
\ee
Denote by $|\Psi\rangle$ the state of the entire quantum
computer, $B=\bbox 1^{(n)}\otimes A$, and consider the following
nonlinear equation 
\be
i|\dot \Psi\rangle &=& \epsilon\tanh \Big(
\frac{\langle\Psi|P^{(n)}|\Psi\rangle}{\langle\Psi|P^{(n)}
B|\Psi\rangle}
-
\frac{\langle\Psi|P^{(n)}P|\Psi\rangle}{\langle
\Psi|P^{(n)}PBP|\Psi\rangle}\Big)P^{(n)} B |\Psi\rangle
\ee
Both expressions occuring under $\tanh$ are time-independent. 
In particular, for $|\Psi\rangle=\sum_{i_1\dots i_ni}\Psi_{i_1\dots i_ni}
|i_1,\dots i_n\rangle|i\rangle$
\be
\frac{\langle\Psi|P^{(n)}P|\Psi\rangle}{\langle
\Psi|P^{(n)}PBP|\Psi\rangle}=
\frac{|\Psi_{0_1\dots 0_n0}|^2}{|\Psi_{0_1\dots 0_n0}|^2\langle
0|A|0\rangle} =\frac{1}{\eta}
\ee
The other term is constant since the operators under the
averages commute with $P^{(n)} B$. The term reads explicitly 
\be
\frac{\langle\Psi|P^{(n)}|\Psi\rangle}{\langle
\Psi|P^{(n)}B|\Psi\rangle}=
\frac{\sum_k|\Psi_{0_1\dots 0_nk}|^2}{\sum_{kl}\Psi^*_{0_1\dots
0_nk} \Psi_{0_1\dots 0_nl}\langle
k|A|l\rangle} 
\ee
We know that 
\be
\Psi_{0_1\dots 0_n0} &=& \frac{2^n-s}{2^n}\\
\Psi_{0_1\dots 0_n1} &=& \frac{s}{2^n}
\ee
and therefore the mobility frequency is identical to the one obtained for a
single qubit description. The explicit evolution of the entire
entangled state of the quantum computer is finally 
\be
|\Psi_t\rangle &=&
\Big(
\bbox 1 -P^{(n)} +P^{(n)}\cos \omega'_\epsilon t -iP^{(n)}B\sin
\omega'_\epsilon t \Big)|\Psi_0\rangle 
\ee
For those of the readers who have played a little bit with
faster-than-light telegraphs in nonlinear quantum mechanics the
basis dependence of the evolution may look somewhat suspicious.
There is no problem with this, however, since the dependence on
$P^{(n)}$ reflects our experimental configuration: By changing
the projector we change the dynamics since we simply put the
nonlinear device in a different position with respect to the
first analyzer. In the faster-than-light problem one gets into
trouble if such basis-dependent terms are produced at a
distance, and this is typical of the Weinberg formulation. 

It may be instructive to discuss what would have happened if we
did not assume that the nonlinearity is somehow activated in a
state dependent way. We therefore assume that the flag system
does not interact with the input one. For this reason we cannot
have any dependence on the basis corresponding to the input particles during
the nonlinear evolution and we use  the Polchinski-type extension of the
dynamics which looks as follows
\be
i|\dot \Psi\rangle &=& \epsilon\tanh \Big(
\frac{\langle\Psi|\Psi\rangle}{\langle\Psi|B|\Psi\rangle}
-
\frac{\langle\Psi|P|\Psi\rangle}{\langle\Psi|PBP|\Psi\rangle}\Big)B
|\Psi\rangle
\ee
The first term under $\tanh$ is obviously time-independent. The
same with the second one which equals
\be
\frac{\langle\Psi|P|\Psi\rangle}{\langle\Psi|PBP|\Psi\rangle}
&=&
\frac{\sum_{i_1\dots i_n}|\Psi_{i_1\dots i_n0}|^2}
{\sum_{i_1\dots i_n}|\Psi_{i_1\dots i_n0}|^2\eta} =\frac{1}{\eta}
\ee
as before. The solution for the entangled state of our quantum
computer is now
\be
|\Psi_t\rangle &=&
\Big(
\bbox 1^{(n+1)} \cos \tilde \omega_\epsilon t -iB\sin
\tilde \omega_\epsilon t \Big)|\Psi_0\rangle 
\ee
where $\tilde \omega_\epsilon$ has to be determined.

To do so we first compute the reduced density matrix of the flag
subsystem
\be
{}&{}&
{\rm Tr}_{1\dots n}|\Psi\rangle \langle\Psi|\nonumber\\
&{}&=
\frac{1}{4^n}\sum_{\vec j}\sum_{\vec i,\vec i'}
(-1)^{(\vec i+\vec i')\cdot \vec j}
|f(i_1,\dots,i_n)\rangle\langle f(i'_1,\dots,i'_n)|\\
&{}&=
\frac{1}{4^n}\sum_{\vec j}\sum_{\vec i}
|f(i_1,\dots,i_n)\rangle\langle f(i_1,\dots,i_n)|\nonumber\\
&{}&+\frac{1}{4^n}\sum_{\vec j}\sum_{\vec i\neq\vec i'}
(-1)^{(\vec i+\vec i')\cdot \vec j}
|f(i_1,\dots,i_n)\rangle\langle f(i'_1,\dots,i'_n)|\\
&{}&=
\frac{2^n-s}{2^n}
|0\rangle\langle 0|+
\frac{s}{2^n}
|1\rangle\langle 1|\nonumber\\
&{}&+\frac{1}{4^n}\sum_{\vec i\neq\vec i'}\sum_{\vec j}
(-1)^{(\vec i+\vec i')\cdot \vec j}
|f(i_1,\dots,i_n)\rangle\langle f(i'_1,\dots,i'_n)|\\
&{}&=
\frac{2^n-s}{2^n}
|0\rangle\langle 0|+
\frac{s}{2^n}
|1\rangle\langle 1|
\ee
because the sums over $\vec j$ vanish. The flag susbsystem is
therefore in a fully mixed state. 
Finally 
\be
\tilde\omega_\epsilon&=&
\epsilon\tanh\Big(\frac{2^n}{(2^n-2s)\eta}-\frac{1}{\eta}\Big)\nonumber\\
&=&
\epsilon\tanh\Big(\frac{s}{(2^{n-1}-s)\eta}\Big)
\approx 
\epsilon\tanh\Big(\frac{s}{2^{n-1}\eta}\Big)\ll \omega'_\epsilon
\ee
so it may pay to act with the nonlinearity on a selected subbeam. 

Returning to the question of exponential precision we should
note that the nonlinearity I have chosen leads to periodic dynamics
and for this reason has a vanishing Lyapunov
exponent. 
One could invent a nonlinear equation for a two-dimensional
dynamics with a positive exponent (cf. \cite{MP}) but calculations
might be less trivial. 

\section{Second algorithm} 

The first three steps are identical to the previous ones.

\medskip
\noindent
{\it Step 4.\/} 
We begin with the result of the third step
\be
|\psi[2]\rangle
&=&
\frac{1}{\sqrt{2^n}}\sum_{i_1\dots i_n=0}^1
|i_1,\dots,i_n\rangle|f(i_1,\dots,i_n)\rangle\label{state2}
\ee
We assume that $f(n)=1$ for at most one $n$ ($s$ equals 0 or 1). 
The state (\ref{state2}) can be written as
\be
|\psi[2]\rangle
&=&
\frac{1}{\sqrt{2^n}}\sum_{i_2\dots i_n=0}^1
|0_1,i_2,\dots,i_n\rangle|f(0_1,i_2,\dots,i_n)\rangle\nonumber\\
&\pp =&+
\frac{1}{\sqrt{2^n}}\sum_{i_2\dots i_n=0}^1
|1_1,i_2,\dots,i_n\rangle|f(1_1,i_2,\dots,i_n)\rangle
\ee
Let us note that with very high probability the state is 
\be
|\psi[2]\rangle
&=&
\frac{1}{\sqrt{2^n}}\sum_{i_2\dots i_n=0}^1
|0_1,i_2,\dots,i_n\rangle|0\rangle\nonumber\\
&\pp =&+
\frac{1}{\sqrt{2^n}}\sum_{i_2\dots i_n=0}^1
|1_1,i_2,\dots,i_n\rangle|0\rangle\label{51}
\ee
With much smaller probability it is either
\be
|\psi[2]\rangle
&=&
\frac{1}{\sqrt{2^n}}\sum_{i_2\dots i_n=0}^1
|0_1,i_2,\dots,i_n\rangle|1\rangle\nonumber\\
&\pp =&+
\frac{1}{\sqrt{2^n}}\sum_{i_2\dots i_n=0}^1
|1_1,i_2,\dots,i_n\rangle|0\rangle\label{52}
\ee
or
\be
|\psi[2]\rangle
&=&
\frac{1}{\sqrt{2^n}}\sum_{i_2\dots i_n=0}^1
|0_1,i_2,\dots,i_n\rangle|0\rangle\nonumber\\
&\pp =&+
\frac{1}{\sqrt{2^n}}\sum_{i_2\dots i_n=0}^1
|1_1,i_2,\dots,i_n\rangle|1\rangle\label{53}
\ee
and is never in the form
\be
|\psi[2]\rangle
&=&
\frac{1}{\sqrt{2^n}}\sum_{i_2\dots i_n=0}^1
|0_1,i_2,\dots,i_n\rangle|1\rangle\nonumber\\
&\pp =&+
\frac{1}{\sqrt{2^n}}\sum_{i_2\dots i_n=0}^1
|1_1,i_2,\dots,i_n\rangle|1\rangle
\ee
since this would mean there are two different numbers satisfying
$f(n)=1$ which contradicts our assumption.
The idea of the algorithm is to use a nonlinearity that leaves
(\ref{51}) unchanged but (\ref{52}) and (\ref{53}) transforms,
respectively, into 
\be
|\psi[2]\rangle
&=&
\frac{1}{\sqrt{2^n}}\sum_{i_2\dots i_n=0}^1
|0_1,i_2,\dots,i_n\rangle|1\rangle\nonumber\\
&\pp =&+
\frac{1}{\sqrt{2^n}}\sum_{i_2\dots i_n=0}^1
|1_1,i_2,\dots,i_n\rangle|1\rangle\label{52'}
\ee
and
\be
|\psi[2]\rangle
&=&
\frac{1}{\sqrt{2^n}}\sum_{i_2\dots i_n=0}^1
|0_1,i_2,\dots,i_n\rangle|1\rangle\nonumber\\
&\pp =&+
\frac{1}{\sqrt{2^n}}\sum_{i_2\dots i_n=0}^1
|1_1,i_2,\dots,i_n\rangle|1\rangle\label{53'}
\ee
One should be aware of the fact such transformations are in fact
impossible within the nonlinear Schr\"dinger equation framework
since one cannot merge two different vectors into a single one
if the dynamics is reversible and first order in time. However
one can do this with arbitrary acuracy as can be clearly seen
from the preceding examples. The more serious problem is that
using exactly the same trick it was shown in \cite{MCfpl} that
this kind of evolution leads to influences between separated
systems (here the flag system would influence the first qubit). 
In the Weinberg description this leads to a contradiction when
one obtains this kind of behavior assuming simultaneously that
the subsystems do not interact. 

We have again two possibilities. We can either assume some form
of interaction between the subsystems, or take a correct
$(n+1)$-particle extension of a nonlinear dynamics of the flag
subsystem assuming no interactions between different
subsystems. 

Before launching into a more detailed analysis 
let us first illustrate the Abrams-Lloyd idea on a simple example.
Take $n=3$ and $f(110)=1$. The oracle produces
\be
{}&{}&\frac{1}{2\sqrt{2}}   \big[ |000\rangle|0\rangle +\nonumber\\
&{}&\pp {\frac{1}{2\sqrt{2}}\big[}|001\rangle|0\rangle +\nonumber\\
&{}&\pp {\frac{1}{2\sqrt{2}}\big[}|010\rangle|0\rangle +\nonumber\\
&{}&\pp {\frac{1}{2\sqrt{2}}\big[}|011\rangle|0\rangle +\nonumber\\
&{}&\pp {\frac{1}{2\sqrt{2}}\big[}|100\rangle|0\rangle +\nonumber\\
&{}&\pp {\frac{1}{2\sqrt{2}}\big[}|101\rangle|0\rangle +\nonumber\\
&{}&\pp {\frac{1}{2\sqrt{2}}\big[}|110\rangle|1\rangle +\nonumber\\
&{}&\pp {\frac{1}{2\sqrt{2}}\big[}|111\rangle|0\rangle \big]
\ee
The nonlinearity now looks at the {\it second\/} and the {\it
third\/} input slots and sees the above kets as the following pairs 
\be
{}&{}&\frac{1}{2\sqrt{2}}   
\big[ |000\rangle|0\rangle +|100\rangle|0\rangle +\nonumber\\
&{}&\pp {\frac{1}{2\sqrt{2}}\big[} 
|001\rangle|0\rangle +|101\rangle|0\rangle +\nonumber\\
&{}&\pp {\frac{1}{2\sqrt{2}}\big[} 
|010\rangle|0\rangle +|110\rangle|1\rangle +\nonumber\\
&{}&\pp {\frac{1}{2\sqrt{2}}\big[} 
|011\rangle|0\rangle +|111\rangle|0\rangle \big]
\ee
Now it scans each of the rows and does not do anything when two
flag 0's occur, but when it notices one 0 and one 1 it changes 0
to 1. So after this step we get 
\be
{}&{}&\frac{1}{2\sqrt{2}}   \big[ 
|000\rangle|0\rangle +|100\rangle|0\rangle +\nonumber\\
&{}&\pp {\frac{1}{2\sqrt{2}}\big[} 
|001\rangle|0\rangle +|101\rangle|0\rangle +\nonumber\\
&{}&\pp {\frac{1}{2\sqrt{2}}\big[} 
|010\rangle|1\rangle +|110\rangle|1\rangle +\nonumber\\
&{}&\pp {\frac{1}{2\sqrt{2}}\big[} 
|011\rangle|0\rangle +|111\rangle|0\rangle \big]
\ee
Now the nonlinearity looks at the {\it first\/} and the {\it
third\/} slots and sees the kets as the following pairs
\be
{}&{}&\frac{1}{2\sqrt{2}}   \big[ 
|000\rangle|0\rangle +|010\rangle|1\rangle +\nonumber\\
&{}&\pp {\frac{1}{2\sqrt{2}}\big[} 
|001\rangle|0\rangle +|011\rangle|0\rangle +\nonumber\\
&{}&\pp {\frac{1}{2\sqrt{2}}\big[} 
|100\rangle|0\rangle +|110\rangle|1\rangle +\nonumber\\
&{}&\pp {\frac{1}{2\sqrt{2}}\big[} 
|101\rangle|0\rangle +|111\rangle|0\rangle \big]
\ee
It again behaves as before and what we get after this step looks
as follows
\be
{}&{}&\frac{1}{2\sqrt{2}} \big[ |000\rangle|1\rangle
+|010\rangle|1\rangle +\nonumber\\ 
&{}&\pp {\frac{1}{2\sqrt{2}}\big[} |001\rangle|0\rangle
+|011\rangle|0\rangle +\nonumber\\ 
&{}&\pp {\frac{1}{2\sqrt{2}}\big[} |100\rangle|1\rangle
+|110\rangle|1\rangle +\nonumber\\ 
&{}&\pp {\frac{1}{2\sqrt{2}}\big[} |101\rangle|0\rangle
+|111\rangle|0\rangle \big] 
\ee
Finally our nonlinearity looks at the {\it first\/} and the {\it
second\/} slots and the state regroups in the following way
\be
{}&{}&\frac{1}{2\sqrt{2}} \big[ |000\rangle|1\rangle
+|001\rangle|0\rangle +\nonumber\\ 
&{}&\pp {\frac{1}{2\sqrt{2}}\big[} |010\rangle|1\rangle
+|011\rangle|0\rangle +\nonumber\\ 
&{}&\pp {\frac{1}{2\sqrt{2}}\big[} |100\rangle|1\rangle
+|101\rangle|0\rangle +\nonumber\\ 
&{}&\pp {\frac{1}{2\sqrt{2}}\big[} |110\rangle|1\rangle
+|111\rangle|0\rangle \big] 
\ee
Now each row contains one 1 and in the final move all flag 0's
are switched to 1's and the state partly disentangles:
\be
{}&{}&\frac{1}{2\sqrt{2}}   \big[ |000\rangle +|001\rangle +\nonumber\\
&{}&\pp {\frac{1}{2\sqrt{2}}\big[} |010\rangle +|011\rangle +\nonumber\\
&{}&\pp {\frac{1}{2\sqrt{2}}\big[} |100\rangle +|101\rangle +\nonumber\\
&{}&\pp {\frac{1}{2\sqrt{2}}\big[} |110\rangle +|111\rangle \big]|1\rangle
\ee
Of course, in case $s=0$ the entire state does not change during
the operation and a measurement on the flag qubit gives 0 with
certainty. 

One can try to implement such an evolution in terms of a
Schr\"odinger-type dynamics. Let us note that the above
procedure is somewhat artificial. Once we agree
that the nonlinearity can somehow globaly and simultaneously
recognize the states of all the qubits  the
optimal strategy would be to choose a nonlinear evolution
which changes all flag 0's into 1's if at least one 1 has been ``seen".

As has been already said this kind of dynamics is unacceptable if
one wants to apply the nonlinear evolution locally only to the flag
qubit. 
Let us proceed therefore differently and apply the
Polchinski-type description. Begin with the nonlinear 1-particle
equation 
\be
i|\dot \psi\rangle &=&
\epsilon\tanh\Big(\alpha\langle\psi|A-\eta\bbox
1|\psi\rangle\Big)A |\psi\rangle
\ee
where $\eta$ and $A$ are the same as before but $\alpha$ is a
very large number. For $|\psi\rangle=|0\rangle$ the expression
under $\tanh$ vanishes. For a small admixture of
$|1\rangle$ and sufficiently large $\alpha$ the mobility
with a nonzero frequency begins and an arbitrary small amount of
$|1\rangle$ can be sufficiently amplified. The extension to the
entire quantum computer is 
\be
i|\dot \Psi\rangle &=&
\epsilon\tanh\Big(\alpha\langle\Psi|\bbox 1^{(n)}\otimes(A-\eta\bbox
1)|\Psi\rangle\Big)\bbox 1^{(n)}\otimes A |\Psi\rangle
\ee
The solution is
\be
|\Psi_t\rangle &=&
\Big(
\bbox 1^{(n+1)} \cos  \omega_{\epsilon,\alpha} t -iB\sin
 \omega_{\epsilon,\alpha} t \Big)|\Psi_0\rangle 
\ee
with 
\be
\omega_{\epsilon,\alpha}
&=&
\epsilon\tanh\Big(\alpha\langle\Psi|\bbox 1^{(n)}\otimes(A-\eta\bbox
1)|\Psi\rangle\Big)\nonumber\\
&=&
\epsilon\tanh\Big(\alpha{\rm Tr\,}\rho(A-\eta\bbox
1)\Big)
\ee
where 
\be
{\rm Tr\,}\rho(A-\eta\bbox 1)
&=&
2^{-n}{\rm Tr}
\left(
\begin{array}{cc}
2^n-s & 0\\
0 & s
\end{array}
\right)
\left(
\begin{array}{cc}
0 & \sqrt{1-\eta^2}\\
\sqrt{1-\eta^2} & -2\eta
\end{array}
\right)=-2^{-n}2\eta s
\ee
so 
\be
\omega_{\epsilon,\alpha}=\epsilon\tanh\Big(\frac{\alpha\eta
s}{2^{n-1}}\Big) 
\ee
Now we can explicitly calculate the average of
$\sigma_3=|0\rangle\langle 0|- |1\rangle\langle 1|$ at the flag
subsystem:
\be
\langle\Psi_t|\bbox 1^{(n)}\otimes\sigma_3|\Psi_t\rangle
&=&
\langle\Psi_0|
\Big(
\bbox 1^{(n+1)} \cos  \omega_{\epsilon,\alpha} t +iB\sin
 \omega_{\epsilon,\alpha} t \Big)\bbox 1^{(n)}\otimes\sigma_3
\Big(
\bbox 1^{(n+1)} \cos  \omega_{\epsilon,\alpha} t -iB\sin
 \omega_{\epsilon,\alpha} t \Big)
|\Psi_0\rangle\nonumber\\
&=&
\langle\Psi_0|
\cos^2  \omega_{\epsilon,\alpha} t 
\bbox 1^{(n)}\otimes\sigma_3
|\Psi_0\rangle\nonumber\\
&\pp =&+
\langle\Psi_0|
\sin^2 \omega_{\epsilon,\alpha} t B\bbox 1^{(n)}\otimes\sigma_3 B
|\Psi_0\rangle\nonumber\\
&\pp =&+i
\langle\Psi_0|
\sin \omega_{\epsilon,\alpha} t 
\cos \omega_{\epsilon,\alpha} t 
B\bbox 1^{(n)}\otimes\sigma_3
|\Psi_0\rangle\nonumber\\
&\pp =&-i
\langle\Psi_0|
\sin \omega_{\epsilon,\alpha} t 
\cos \omega_{\epsilon,\alpha} t 
\bbox 1^{(n)}\otimes\sigma_3 B
|\Psi_0\rangle\nonumber\\
&=&
\cos^2  \omega_{\epsilon,\alpha} t 
\frac{2^n-2s}{2^n}
+
\sin^2 \omega_{\epsilon,\alpha} t
\langle\Psi_0|
\bbox 1^{(n)}\otimes A\sigma_3 A
|\Psi_0\rangle\nonumber\\
&\pp =&+\frac{i}{2}\sin 2\omega_{\epsilon,\alpha} t
\langle\Psi_0|
\bbox 1^{(n)}\otimes[A,\sigma_3]
|\Psi_0\rangle\nonumber\\
&=&
\cos^2  \omega_{\epsilon,\alpha} t 
\frac{2^{n-1}-s}{2^{n-1}}
+
\sin^2 \omega_{\epsilon,\alpha} t
\langle\Psi_0|
\bbox 1^{(n)}\otimes A\sigma_3 A
|\Psi_0\rangle\nonumber\\
&\pp =&+\frac{i}{2}\sqrt{1-\eta^2}\sin 2\omega_{\epsilon,\alpha} t
\langle\Psi_0|
\bbox 1^{(n)}\otimes[\sigma_1,\sigma_3]
|\Psi_0\rangle
\ee
Now
\be
{\rm Tr\,}\rho A\sigma_3 A 
&=&
\frac{1}{2^n}
{\rm Tr\,}\left(
\begin{array}{cc}
2^n-s & 0\\
0 & s
\end{array}
\right)
\left(
\begin{array}{cc}
\eta & \sqrt{1-\eta^2}\\
\sqrt{1-\eta^2} & -\eta
\end{array}
\right)
\left(
\begin{array}{cc}
1 & 0\\
0 & -1
\end{array}
\right)
\left(
\begin{array}{cc}
\eta & \sqrt{1-\eta^2}\\
\sqrt{1-\eta^2} & -\eta
\end{array}
\right)\nonumber\\
&=&
\frac{1}{2^n}
{\rm Tr\,}\left(
\begin{array}{cc}
2^n-s & 0\\
0 & s
\end{array}
\right)
\left(
\begin{array}{cc}
\eta & -\sqrt{1-\eta^2}\\
\sqrt{1-\eta^2} & \eta
\end{array}
\right)
\left(
\begin{array}{cc}
\eta & \sqrt{1-\eta^2}\\
\sqrt{1-\eta^2} & -\eta
\end{array}
\right)\nonumber\\
&=&
\frac{1}{2^n}
{\rm Tr\,}\left(
\begin{array}{cc}
2^n-s & 0\\
0 & s
\end{array}
\right)
\left(
\begin{array}{cc}
2\eta^2-1 & 2\eta\sqrt{1-\eta^2}\\
2\eta\sqrt{1-\eta^2} & 1-2\eta^2
\end{array}
\right)\nonumber\\
&=&
\frac{1}{2^n}(2^n-2s)(2\eta^2-1)
\ee
\be
\langle\Psi_t|\bbox 1^{(n)}\otimes\sigma_3|\Psi_t\rangle
&=&
\frac{2^{n-1}-s}{2^{n-1}}\cos^2  \omega_{\epsilon,\alpha} t 
+
\frac{(2^{n-1}-s)(2\eta^2-1)}{2^{n-1}}
\sin^2 \omega_{\epsilon,\alpha} t
\nonumber\\
&=&
\frac{2^{n-1}-s}{2^{n-1}}\cos 2 \omega_{\epsilon,\alpha} t 
+
2\eta^2\frac{2^{n-1}-s}{2^{n-1}}\sin^2 \omega_{\epsilon,\alpha} t
\ee
For $s=0$ the average is constant in time and equals 1. For
$s\neq 0$ and $\eta^2\approx 0$ and sufficiently large $\alpha$
it oscillates with $\omega_{\epsilon,\alpha}\approx\epsilon$.
This kind of algorithm cannot distinguish between different
nonzero values of $s$, but clearly distinguishes between $s=0$
and $s\neq 0$ in a way that is insensitive to small fluctuations of
the parameters.

What is important, such an algorithm is obtained by applying the
nolinear evolution only to the flag qubit. This is done in a
fully local way. We conclude that nonlinear quantum evolutions
can lead to fast algorithms and the fact that they are fast does
not follow from unphysical faster-than-light effects.


\begin{references}
\bibitem{AL}D. S. Abrams and S. Lloyd, 
``Nonlinear quantum mechanics implies polynomial-time solution for
NP-complete and \#P problems", quant-ph (January 1998) 
\bibitem{MCfpl}M.~Czachor, ``Mobility and nonseparability",
Found. Phys. Lett. {\bf 4}, 351 (1991).
\bibitem{Mielnik}B. Mielnik, ``Mobility of nonlinear systems",
J. Math. Phys. {\bf 21}, 44 (1980).
\bibitem{MP}D. Makowiec and A. Posiewnik, ``Analysis of
nonlinear Schr\"odinger equation for spin", Phys. Lett. A {\bf
207}, 37 (1995).
\bibitem{MCqph}M.~Czachor, ``Nonlocally looking equations can
make nonlinear quantum dynamics local",
Report No. quant-ph/9708052.
\bibitem{MCpra}M.~Czachor, ``Nonlinear Schr\"odinger equation
and two-level atoms", Phys.~Rev.~A {\bf 53}, 1310 (1996).
\end{references}
\end{document}